\PassOptionsToPackage{unicode}{hyperref}
\PassOptionsToPackage{hyphens}{url}
\PassOptionsToPackage{dvipsnames,svgnames,x11names}{xcolor}
\documentclass[
  11pt,
  letterpaper,
  DIV=11,
  numbers=noendperiod]{scrartcl}

\usepackage{amsmath,amssymb}
\usepackage{iftex}
\ifPDFTeX
  \usepackage[T1]{fontenc}
  \usepackage[utf8]{inputenc}
  \usepackage{textcomp} 
\else 
  \usepackage{unicode-math}
  \defaultfontfeatures{Scale=MatchLowercase}
  \defaultfontfeatures[\rmfamily]{Ligatures=TeX,Scale=1}
\fi
\usepackage{lmodern}
\ifPDFTeX\else  
\fi
\IfFileExists{upquote.sty}{\usepackage{upquote}}{}
\IfFileExists{microtype.sty}{
  \usepackage[]{microtype}
  \UseMicrotypeSet[protrusion]{basicmath} 
}{}
\makeatletter
\@ifundefined{KOMAClassName}{
  \IfFileExists{parskip.sty}{%
    \usepackage{parskip}
  }{
    \setlength{\parindent}{0pt}
    \setlength{\parskip}{6pt plus 2pt minus 1pt}}
}{
  \KOMAoptions{parskip=half}}
\makeatother
\usepackage{xcolor}
\usepackage[lmargin=1in,rmargin=1in,tmargin=1in,bmargin=1in]{geometry}
\makeatletter
\ifx\paragraph\undefined\else
  \let\oldparagraph\paragraph
  \renewcommand{\paragraph}{
    \@ifstar
      \xxxParagraphStar
      \xxxParagraphNoStar
  }
  \newcommand{\xxxParagraphStar}[1]{\oldparagraph*{#1}\mbox{}}
  \newcommand{\xxxParagraphNoStar}[1]{\oldparagraph{#1}\mbox{}}
\fi
\ifx\subparagraph\undefined\else
  \let\oldsubparagraph\subparagraph
  \renewcommand{\subparagraph}{
    \@ifstar
      \xxxSubParagraphStar
      \xxxSubParagraphNoStar
  }
  \newcommand{\xxxSubParagraphStar}[1]{\oldsubparagraph*{#1}\mbox{}}
  \newcommand{\xxxSubParagraphNoStar}[1]{\oldsubparagraph{#1}\mbox{}}
\fi
\makeatother

\usepackage{color}
\usepackage{fancyvrb}

\DefineVerbatimEnvironment{Highlighting}{Verbatim}{commandchars=\\\{\}}
\usepackage{framed}
\definecolor{shadecolor}{RGB}{241,243,245}
\newenvironment{Shaded}{\begin{snugshade}}{\end{snugshade}}

\newcommand{\AttributeTok}[1]{\textcolor[rgb]{0.40,0.45,0.13}{#1}}

\newcommand{\CharTok}[1]{\textcolor[rgb]{0.13,0.47,0.30}{#1}}

\newcommand{\DataTypeTok}[1]{\textcolor[rgb]{0.68,0.00,0.00}{#1}}
\newcommand{\DecValTok}[1]{\textcolor[rgb]{0.68,0.00,0.00}{#1}}

\newcommand{\FunctionTok}[1]{\textcolor[rgb]{0.28,0.35,0.67}{#1}}
\newcommand{\ImportTok}[1]{\textcolor[rgb]{0.00,0.46,0.62}{#1}}

\newcommand{\KeywordTok}[1]{\textcolor[rgb]{0.00,0.23,0.31}{\textbf{#1}}}
\newcommand{\NormalTok}[1]{\textcolor[rgb]{0.00,0.23,0.31}{#1}}
\newcommand{\OperatorTok}[1]{\textcolor[rgb]{0.37,0.37,0.37}{#1}}

\newcommand{\StringTok}[1]{\textcolor[rgb]{0.13,0.47,0.30}{#1}}

\providecommand{\tightlist}{%
  \setlength{\itemsep}{0pt}\setlength{\parskip}{0pt}}\usepackage{longtable,booktabs,array}
\usepackage{calc} 
\usepackage{etoolbox}
\makeatletter
\patchcmd\longtable{\par}{\if@noskipsec\mbox{}\fi\par}{}{}
\makeatother
\IfFileExists{footnotehyper.sty}{\usepackage{footnotehyper}}{\usepackage{footnote}}
\makesavenoteenv{longtable}
\usepackage{graphicx}
\makeatletter
\newsavebox\pandoc@box
\newcommand*\pandocbounded[1]{
  \sbox\pandoc@box{#1}%
  \Gscale@div\@tempa{\textheight}{\dimexpr\ht\pandoc@box+\dp\pandoc@box\relax}%
  \Gscale@div\@tempb{\linewidth}{\wd\pandoc@box}%
  \ifdim\@tempb\p@<\@tempa\p@\let\@tempa\@tempb\fi
  \ifdim\@tempa\p@<\p@\scalebox{\@tempa}{\usebox\pandoc@box}%
  \else\usebox{\pandoc@box}%
  \fi%
}
\def\fps@figure{htbp}
\makeatother
\NewDocumentCommand\citeproctext{}{}

\makeatletter
 \let\@cite@ofmt\@firstofone
 \def\@biblabel#1{}
 \def\@cite#1#2{{#1\if@tempswa , #2\fi}}
\makeatother
\newlength{\cslhangindent}
\newlength{\csllabelwidth}
\newenvironment{CSLReferences}[2] 
 {\begin{list}{}{%
  \setlength{\itemindent}{0pt}
  \setlength{\leftmargin}{0pt}
  \setlength{\parsep}{0pt}
  \ifodd #1
   \setlength{\leftmargin}{\cslhangindent}
   \setlength{\itemindent}{-1\cslhangindent}
  \fi
  \setlength{\itemsep}{#2\baselineskip}}}
 {\end{list}}
\usepackage{calc}

\KOMAoption{captions}{tableheading}
\makeatletter
\@ifpackageloaded{caption}{}{\usepackage{caption}}
\AtBeginDocument{%
\ifdefined\contentsname
  \renewcommand*\contentsname{Table of contents}
\else
  \newcommand\contentsname{Table of contents}
\fi
\ifdefined\listfigurename
  \renewcommand*\listfigurename{List of Figures}
\else
  \newcommand\listfigurename{List of Figures}
\fi
\ifdefined\listtablename
  \renewcommand*\listtablename{List of Tables}
\else
  \newcommand\listtablename{List of Tables}
\fi
\ifdefined\figurename
  \renewcommand*\figurename{Figure}
\else
  \newcommand\figurename{Figure}
\fi
\ifdefined\tablename
  \renewcommand*\tablename{Table}
\else
  \newcommand\tablename{Table}
\fi
}
\@ifpackageloaded{float}{}{\usepackage{float}}
\floatstyle{ruled}
\@ifundefined{c@chapter}{\newfloat{codelisting}{h}{lop}}{\newfloat{codelisting}{h}{lop}[chapter]}
\floatname{codelisting}{Listing}

\makeatother
\makeatletter
\@ifpackageloaded{caption}{}{\usepackage{caption}}
\@ifpackageloaded{subcaption}{}{\usepackage{subcaption}}
\makeatother

\usepackage{bookmark}

\IfFileExists{xurl.sty}{\usepackage{xurl}}{} 
\hypersetup{
  pdftitle={OASIS: A Rubric-Based Multimodal Assessment Platform Using Large Language Models},
  pdfauthor={Ameer H. Shakur\^{}\{1,\textbackslash dagger\}; Shinyoung Kang\^{}\{1,\textbackslash dagger\}; David Hein\^{}\{1,\textbackslash dagger\}; Michael J. Holcomb\^{}1; Huong-Tra Ngo\^{}1; Aarash Zakeri\^{}3; Minhan Park\^{}1; Licheng Yi\^{}\{1,5\}; Dhanush Jain\^{}4; Hunter Schuler\^{}2; Andrew R. Jamieson\^{}\{1,*\}},
  colorlinks=true,
  linkcolor={black},
  filecolor={Maroon},
  citecolor={blue},
  urlcolor={blue},
  pdfcreator={LaTeX via pandoc}}

\title{OASIS: A Rubric-Based Multimodal Assessment Platform Using Large
Language Models}
\usepackage{etoolbox}
\makeatletter
\providecommand{\subtitle}[1]{
  \apptocmd{\@title}{\par {\large #1 \par}}{}{}
}
\makeatother
\subtitle{Technical Report}
\author{Ameer H. Shakur\(^{1,\dagger}\) \and Shinyoung
Kang\(^{1,\dagger}\) \and David Hein\(^{1,\dagger}\) \and Michael J.
Holcomb\(^1\) \and Huong-Tra Ngo\(^1\) \and Aarash
Zakeri\(^3\) \and Minhan Park\(^1\) \and Licheng
Yi\(^{1,5}\) \and Dhanush Jain\(^4\) \and Hunter
Schuler\(^2\) \and Andrew R. Jamieson\(^{1,*}\)}
\date{August 3, 2026}

\begin{document}
\maketitle
\begin{abstract}
OASIS (Open Assessment and Scoring Infrastructure Stack) is a systems
platform for rubric-based grading of video, audio, and text with large
language models. Scoring one artifact with an LLM is straightforward;
deploying assessment at scale requires encounter management, rubric
versioning, modality-aware execution, provenance capture, and human
review. OASIS pairs a standalone command-line interface with a canonical
integrated Elephant + MAPLES stack for encounter management and
multimodal grading orchestration. Both paths can target hosted APIs or
self-hosted open-weight models through Ollama and OpenAI-compatible
endpoints such as vLLM. SimRubrics rubric authoring and the Wayfinder
conversational agent gateway are optional extensions that use the same
authenticated interfaces as human operators. Given a rubric and recorded
encounters, OASIS produces per-criterion scores, evidence, and
rationales, preserving execution artifacts for audit. Distinctive
features include rubric-as-program compilation, progressive execution
plans, content-addressable grading identity, transcript-augmented
multimodal grading, explicit review state, and a shared command surface
for humans and autonomous agents. Though developed in medical education,
the architecture is domain-agnostic, applying wherever structured
performance can be evaluated from recorded or written artifacts. In
production at UT Southwestern Medical Center since Fall 2023, the
platform has processed more than 7,000 encounters. This publication
includes the report and project information, not application source,
binaries, installation materials, sample data, or a tagged software
release.
\end{abstract}

\vspace{0.5em}
\begin{center}
{\small\itshape $^1$Jamieson Lab, Lyda Hill Department of Bioinformatics, UT Southwestern Medical Center, Dallas, TX, USA\\
$^2$Southern Methodist University, Dallas, TX, USA\\
$^3$University of Pittsburgh, Pittsburgh, PA, USA \quad $^4$University of Texas at Austin, Austin, TX, USA\\
$^5$Texas A\&M University, College Station, TX, USA\\[0.5em]
$^\dagger$Co-lead developers \quad $^*$Corresponding author}
\end{center}
\vspace{1em}

\section{Motivation}\label{motivation}

Structured performance assessments are everywhere --- medical licensing
exams, teacher practicums, language proficiency tests, simulation-based
training. The workflow is repetitive and labor-intensive: an evaluator
watches a recording or reads a written artifact, consults a rubric, and
assigns scores. Cost grows linearly with the number of encounters, while
reliability remains imperfect; human inter-rater reliability on
performance assessments is itself often modest (Downing 2004).

Throughout this report, an \emph{encounter} is any performance artifact
that can be evaluated against a rubric. The term comes from medical
education, but the abstraction is broader: interviews, simulations,
classroom observations, drills, inspections, or structured written
responses. The core vocabulary is small:

\begin{longtable}[]{@{}
  >{\raggedright\arraybackslash}p{(\linewidth - 2\tabcolsep) * \real{0.4000}}
  >{\raggedright\arraybackslash}p{(\linewidth - 2\tabcolsep) * \real{0.6000}}@{}}
\toprule\noalign{}
\begin{minipage}[b]{\linewidth}\raggedright
Term
\end{minipage} & \begin{minipage}[b]{\linewidth}\raggedright
Meaning
\end{minipage} \\
\midrule\noalign{}
\endhead
\bottomrule\noalign{}
\endlastfoot
\textbf{Encounter} & One evaluable performance artifact (or set of
artifacts): a recording, a written response, or both \\
\textbf{Rubric / item} & The evaluator-authored specification; each
\emph{item} is one scorable criterion with leveled descriptors \\
\textbf{Station} & One distinct scoring context within an assessment (an
exam room, a scenario, an inspection setting) \\
\textbf{Modality} & The input type an item is scored from: video, audio,
or text \\
\textbf{Group} & A named collection of encounters selected for grading
together \\
\textbf{Run} & One execution of a rubric against a group (or local file
set), with its own provenance record \\
\textbf{Review state} & The per-item human adjudication status:
\texttt{pending}, \texttt{accepted}, or \texttt{reviewed} (overridden by
a human) \\
\textbf{Execution plan} & The inspectable staged artifact (plan, then
sample, then scale) that gates cost and structural validity \\
\end{longtable}

Recent LLM work has shown that models can score notes, transcripts, and
videos with useful agreement with human raters (Jamieson et al. 2024;
Shakur et al. 2024; Holcomb, Kang, and Shakur 2024). But a single
successful model call is not yet a deployable assessment system. Real
use requires infrastructure to manage encounter data, preserve rubric
and model state, route media to modality-appropriate grading paths,
capture execution artifacts, and support human review before results are
treated as final.

OASIS addresses this systems problem. It is designed for academic
deployment so that institutions can evaluate rubric-based AI assessment
without depending on a proprietary end-to-end vendor, while preserving
the operational properties that matter in practice: traceability,
repeatability, and reviewability. The design thesis throughout is that
LLM assessment becomes trustworthy when it becomes \emph{inspectable
software}: rubrics compile to typed programs, every grade has a stable
content-addressed identity, and human review is a recorded state
transition rather than a spreadsheet ritual. Later sections make this
concrete by following a single rubric item --- \texttt{organization},
from the starter pack's Documentation Quality rubric --- from
spreadsheet row to compiled prompt, typed result, cache identity, review
decision, and replay bundle.

The intended standalone evaluation profile is designed around a single
Go binary and one provider credential. The evaluated command sequence
moves from environment checks to a no-spend plan and then a one-item
sample before any larger run. No binary or installable source is
included in this public publication. An internal synthetic starter pack
was used for first-run validation; it is not included here. The
integrated stack and agent surfaces build on the same workflow, so a
local validation can carry into larger deployments when the software is
available under separately stated terms.

\section{Related Work}\label{related-work}

Prior work falls into three broad categories. First, essay and
short-answer scoring systems address text-only artifacts (Shermis and
Burstein 2013). Second, commercial products such as HireVue and Vervoe
assess interview-style video tasks (HireVue, Inc. 2026; Vervoe Pty Ltd
2026), but they are proprietary, and their publicly documented
interfaces do not expose a general, user-defined rubric workflow. Third,
recent medical-education studies have demonstrated LLM-based scoring of
notes, transcripts, and videos (Jamieson et al. 2024; Shakur et al.
2024; Holcomb, Kang, and Shakur 2024), but the released methods are
typically scoped to individual study pipelines rather than reusable
platform infrastructure.

A parallel line of work studies LLMs as judges of model or human output
(Zheng et al. 2023; Liu et al. 2023). Its documented judge biases ---
position, verbosity, self-preference --- reinforce a central OASIS
premise: LLM-assigned scores need structured human review and complete
provenance before they are treated as final. Model-evaluation harnesses
in that tradition benchmark models against fixed task suites; OASIS
inverts the direction, scoring human performance artifacts against
evaluator-defined rubrics.

General-purpose LLM APIs and workflow orchestrators also solve only part
of the problem. An API can score one prompt, and orchestration tools
such as Airflow (Apache Software Foundation 2015) or Prefect (Prefect
Technologies, Inc. 2024) can schedule tasks, but neither provides the
missing assessment layer: rubric-aware prompt compilation,
modality-aware routing, encounter and artifact management, provenance
from rubric version through model output to final score, and a review
surface for evaluator override.

What is missing, therefore, is not another prompt template or another
workflow engine, but a rubric-native, multimodal assessment system that
integrates data management, execution, provenance, and review in one
architecture. OASIS is intended to fill that gap.

\section{Contributions}\label{contributions}

This technical report makes five contributions:

\begin{enumerate}
\def\labelenumi{\arabic{enumi}.}
\tightlist
\item
  An end-to-end assessment architecture with contract-enforced component
  boundaries, spanning standalone CLI use, institutional deployment, and
  agent-facing operation --- packaged as an umbrella workspace with
  canonical service submodules.
\item
  A rubric-as-program compilation strategy that compiles spreadsheet
  rubrics into modality-aware prompts and typed result schemas.
\item
  A staged multimodal execution model that pairs progressive,
  cost-visible execution plans and transcript-augmented media grading
  with content-addressable grading identity, enabling cache reuse,
  cross-model comparison, and distributed batch execution.
\item
  An agent-native command surface in which CLI, terminal UI, and Model
  Context Protocol (MCP) tools share one command-service layer,
  complemented at the institutional tier by a platform agent gateway
  that exposes the grading workflow as typed MCP tools and embeds a
  skills-driven Wayfinder assistant in the web interface.
\item
  A provider-neutral experimentation surface in which hosted APIs,
  Ollama, and OpenAI-compatible endpoints such as vLLM can execute
  against the same rubric snapshots, content-addressed inputs, review
  workflow, and model-identified result records.
\end{enumerate}

Operational experience --- more than 7,000 production encounters since
Fall 2023, with human review and provenance capture throughout ---
serves as the supporting evidence for these design choices rather than
as a separate contribution.

\section{Evolution and Design
Philosophy}\label{evolution-and-design-philosophy}

OASIS evolved through four overlapping stages: a grading pipeline, a
reusable platform, an agent-native operating surface, and an accessible
tool for non-specialist users. The original MAPLES pipeline established
that LLMs could apply rubrics to clinical performance artifacts at
operational scale. The platform layer added Elephant for encounter and
file management, SimRubrics for rubric quality assurance, and shared
contracts across services.

The agent-native layer then exposed the same workflow through the CLI,
the interactive terminal UI (TUI), and MCP surfaces (Anthropic 2024), so
autonomous assistants could plan, inspect, and recover grading runs
without bypassing the human-facing system. A platform agent gateway and
an embedded Wayfinder chat assistant later brought the same property to
the institutional web interface. (Wayfinder is the platform's assistant
persona; it has two implementations --- an agent embedded in the CLI and
TUI, and a gateway-backed chat in the MAPLES web application --- and
this report qualifies each use as CLI-tier or platform-tier.) The fourth
layer is accessibility: a standalone binary, guided workflow,
reproducibility bundle, and synthetic starter pack let researchers
evaluate the system without deploying institutional infrastructure.

\section{Software Design}\label{software-design}

\subsection{Distinctive System
Features}\label{distinctive-system-features}

OASIS's distinguishing contribution is not a single model prompt or one
service boundary. It is the combination of assessment-specific
contracts, staged execution, and reviewable evidence that turns LLM
scoring into an inspectable software system.

\begin{itemize}
\tightlist
\item
  \textbf{Rubric-as-program compilation.} Spreadsheet criteria become
  modality-aware prompt batches and typed result schemas rather than
  informal prompt text.
\item
  \textbf{Perception-before-assessment.} Text extraction, transcript
  generation, and media routing are treated as explicit artifacts before
  the LLM assigns a score.
\item
  \textbf{Progressive execution plans.} Users and agents can validate
  data, estimate cost, sample, and resume before scaling to a full
  cohort.
\item
  \textbf{Content-addressable grading identity.} Grading units are keyed
  by rubric, input, prompt, and model identity, enabling cache reuse,
  model comparison, and distributed batch work.
\item
  \textbf{Human review as workflow state.} Machine scores remain
  provisional until accepted or overridden, and review evidence links
  back into the run record.
\item
  \textbf{Agent-native operation.} CLI, TUI, and MCP tools share the
  same command-service layer, so agents operate the same workflow as
  humans rather than a separate wrapper; at the institutional tier, a
  platform agent gateway and embedded Wayfinder chat extend the same
  property to the web application.
\item
  \textbf{Provenance and replay bundles.} Each run preserves rubric
  snapshots, prompts, provider/model metadata, outputs, evidence
  indexes, and replay instructions.
\item
  \textbf{Hosted and open-weight model paths.} Ollama and
  OpenAI-compatible endpoints such as vLLM let self-hosted models
  participate in the same grading, comparison, review, and provenance
  workflow as hosted APIs.
\item
  \textbf{Deployment gradient.} The same rubric and artifact model
  supports local standalone grading, the integrated Elephant + MAPLES
  stack, and optional agent-driven orchestration; shared hosting is a
  separate site-owned adapter.
\end{itemize}

\subsection{Architecture}\label{architecture}

\begin{figure}

\centering{

\pandocbounded{\includegraphics[keepaspectratio]{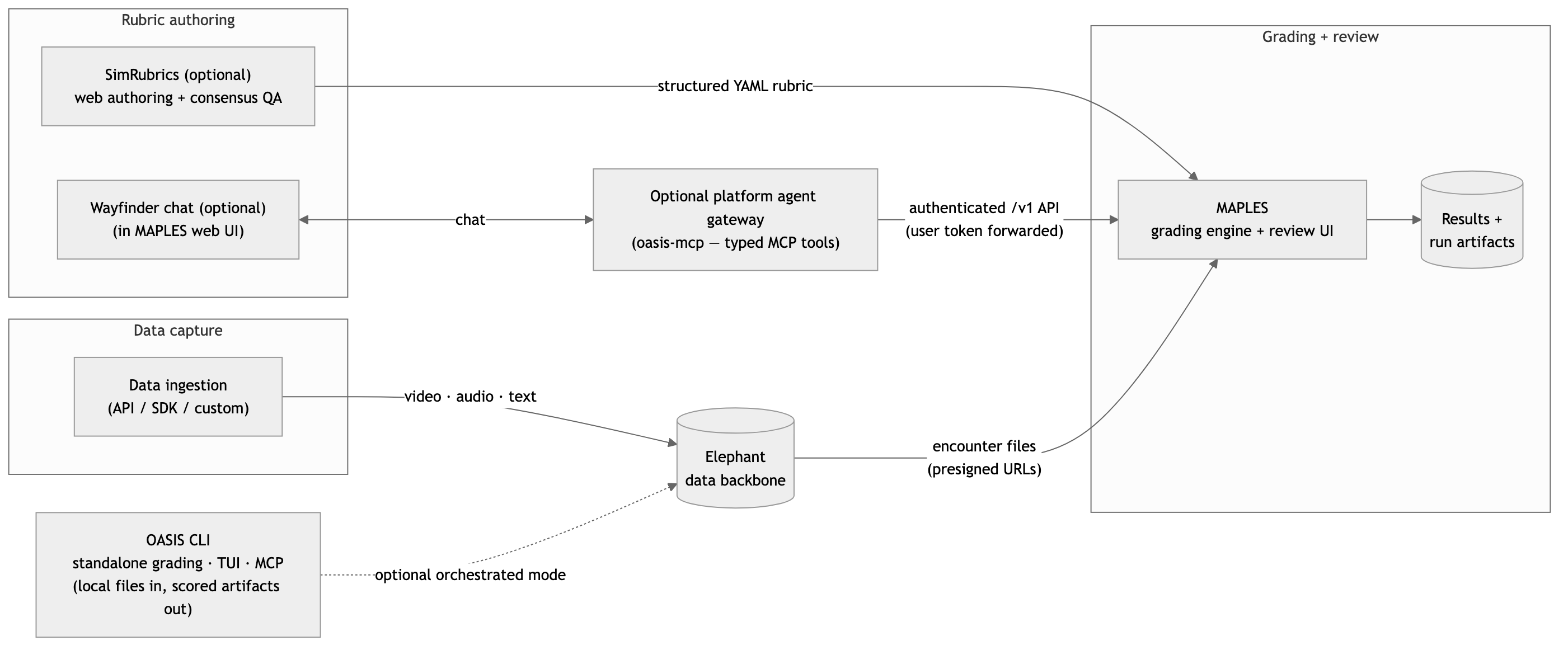}}

}

\caption{\label{fig-architecture}OASIS platform architecture. The
standalone CLI and integrated Elephant + MAPLES path are canonical.
Optional SimRubrics authoring and Wayfinder chat can supply rubrics
through the optional agent gateway to MAPLES's authenticated API. Data
ingestion populates Elephant; MAPLES retrieves encounter files via
presigned URLs, executes grading, and persists results and run
artifacts.}

\end{figure}

OASIS combines a standalone CLI with the integrated Elephant + MAPLES
stack, all organized around a shared rubric and artifact model.
SimRubrics and the platform agent gateway are optional extensions
(Figure~\ref{fig-architecture}).

The standalone CLI reuses the same rubric logic and grading concepts
against local files, but does not require Elephant, MAPLES, or
SimRubrics to be running. Across all five components, code-generated
contracts (OpenAPI specifications, sqlc queries, and Pydantic schemas)
bind services at compile time rather than at runtime.

\textbf{SimRubrics} (Flask, PostgreSQL, Vertex AI). A web application
for AI-assisted rubric quality analysis. Evaluators upload rubrics in
Excel or CSV format; the system analyzes each criterion for clarity and
measurability using multiple LLM providers and exports a structured YAML
file that MAPLES consumes directly. SimRubrics is optional --- users can
also provide rubrics to MAPLES as Excel files without it (see The
Rubric-to-Prompt Transformation).

\textbf{Elephant} (Go, PostgreSQL, MinIO). A REST API for managing
encounter data. It stores metadata (who, what, when, where) in
PostgreSQL and files (video, audio, text) in S3-compatible object
storage. The API is generated from an OpenAPI 3.0 specification; the
Python SDK is generated from the same spec. Files are never proxied ---
clients get presigned URLs and upload/download directly from object
storage. Deduplication and key design are detailed in the Elephant
section below.

\textbf{MAPLES} (Multimodal Assessment Pipeline for Learning Encounter
Scoring; FastAPI, Prefect 3, PostgreSQL, MinIO). The grading engine.
Users select encounters from Elephant, pair them with a rubric, and
submit a pipeline run. Prefect orchestrates parallel grading: each
encounter is processed concurrently, and within each encounter, video,
audio, and text items can be graded in parallel. Video grading uses
Gemini's native multimodal API. Audio is graded natively by capable
providers or routed transcript-first --- transcribed once, then graded
as text. Text follows the same rubric-driven prompt pathway over textual
inputs. Results are stored per criterion with the LLM's rationale, and a
review interface lets evaluators inspect, override, and export results.
MAPLES's three core technical contributions --- the rubric-to-prompt
compiler, the default two-stage grading path, and the flow-scoped
caching strategy --- are detailed below.

\textbf{OASIS CLI} (Go). A single binary providing command-line access
to the grading workflow without requiring server infrastructure. In
standalone mode, the CLI reads rubrics and encounter files from the
local filesystem, calls the LLM API directly, and writes scored results
(CSV, Markdown, and a reproducibility bundle with provenance chain) to
an output directory. This eliminates Docker, PostgreSQL, and MinIO
dependencies for individual researchers or evaluators trying the
platform for the first time. The CLI supports Gemini, OpenAI, Anthropic,
Ollama, and OpenAI-compatible endpoints. It also provides an interactive
terminal UI, multi-model comparison, rubric conversion and editing
utilities, a content-addressable cache that avoids redundant LLM calls
across reruns, and a progressive pipeline that separates planning from
sampling and full execution. An MCP server exposes a generated tool
catalog over JSON-RPC, enabling integration with AI assistants and
autonomous agents.

\textbf{OASIS MCP Gateway (optional)} (Python, FastMCP; the
\emph{platform agent gateway} hereafter). This opt-in counterpart to the
CLI's agent surface exposes the MAPLES REST workflow --- authentication,
rubrics, groups, station mappings, pipelines, results, review,
transcription, and video splitting --- as a catalog of typed MCP tools,
and hosts the chat loop behind the platform-tier Wayfinder assistant
embedded in the MAPLES web interface. Authentication remains with
MAPLES, which forwards the signed-in user's token; the trust boundary
and the skills-driven rubric-authoring workflow are detailed in the
Agent-Native Command Surface section.

\subsection{Repository Topology and Planned
Distribution}\label{repository-topology-and-planned-distribution}

The public
\href{https://github.com/JamiesonLabUTSW/oasis}{\texttt{JamiesonLabUTSW/oasis}}
repository currently carries project information and this technical
report only. The working implementation remains private. A later
software release may add the standalone OASIS CLI (\texttt{oasis-go/}),
the platform agent gateway (\texttt{oasis-mcp/}), sample data,
integration harnesses, and canonical service references, but that future
publication will identify its exact contents, version, and terms. The
intended topology is deliberately not a monolithic code dump: Elephant,
MAPLES, SimRubrics, and the MAPLES Toolkit retain independent component
boundaries while an application workspace records an integrated
snapshot.

This topology encodes architectural boundaries in the repository itself.
Elephant owns contract-first encounter management; MAPLES owns runtime
orchestration and review; SimRubrics owns upstream rubric refinement;
the Toolkit owns agent skill definitions; the CLI owns the low-friction
entry path plus terminal and agent surfaces; and the gateway owns
platform-tier agent access. Shared behavior crosses those boundaries
only through explicit contracts: OpenAPI in the Elephant path, typed
rubric schemas in the grading path, and versioned run artifacts at
execution time.

The separation also improves engineering velocity. Component teams can
evolve their service without turning the umbrella workspace into a large
coupled monolith, while the application repository can add end-to-end
tests, scenario packs, and release gates that exercise the integrated
stack as users actually experience it.

\subsection{Design Decisions}\label{design-decisions}

The system is organized around seven design decisions.

\begin{enumerate}
\def\labelenumi{\arabic{enumi}.}
\tightlist
\item
  \textbf{Rubric-driven.} The evaluator controls what is scored and how.
  The rubric is the specification; the LLM is the executor.
\item
  \textbf{Domain-agnostic.} Nothing in the rubric format or grading
  pipeline is specific to medicine, education, or any other field. If an
  assessment produces video, audio, or text and can be described by a
  rubric, OASIS can process it. Production evidence to date comes from
  medical education; an internal synthetic validation corpus includes a
  non-medical example to keep the abstraction concrete.
\item
  \textbf{Perception before assessment.} The architecture separates
  deterministic perception --- text extraction, audio transcription,
  video frame processing --- from stochastic assessment by the LLM.
  Perception artifacts are cached, validated, and reused across models.
  The principle is that a model should never score what it cannot
  perceive; evidence must be grounded in what was actually extracted
  from the input, not inferred.
\item
  \textbf{Provider-aware, not provider-locked.} The CLI supports Gemini,
  OpenAI, Anthropic, Ollama, and OpenAI-compatible endpoints. MAPLES
  supports Gemini, OpenAI, Azure OpenAI, and OpenAI-compatible endpoints
  --- including locally hosted open-weight servers --- with per-user
  bring-your-own-key credentials; SimRubrics exposes hosted providers
  where multimodal support or hosted integrations matter. The overall
  architecture avoids coupling the workflow to a single vendor.
\item
  \textbf{Cost-visible.} Planning and sampling are first-class stages.
  Users can estimate cost, grade a small subset, and only then scale to
  a full cohort.
\item
  \textbf{Auditable and replayable.} OASIS snapshots rubrics, records
  provider and model identifiers, persists prompts and outputs, and
  captures review actions so that prior runs can be inspected and
  replayed within the limits of external model versioning.
\item
  \textbf{Incremental and human-reviewed.} Users can start with the
  standalone CLI, then add Elephant, MAPLES, and SimRubrics as their
  operational needs grow. AI outputs are not considered final until they
  pass through a human review workflow.
\end{enumerate}

\subsection{Hosted and Open-Weight
Inference}\label{hosted-and-open-weight-inference}

OASIS treats the model endpoint as an explicit execution choice rather
than a fixed platform dependency. The standalone CLI supports Gemini,
OpenAI, Anthropic, Ollama, and OpenAI-compatible endpoints. MAPLES
supports Gemini, OpenAI, Azure OpenAI, and OpenAI-compatible endpoints,
including models served locally with vLLM. A model comparison can
therefore vary the provider, endpoint, or model while retaining the same
rubric snapshot, encounter selection, review states, and result
structure.

To assess compatibility with self-hosted inference, we evaluated
representative models from the Gemma and Qwen families through vLLM in
several illustrative configurations. Examples included direct grading of
text; transcript-mediated audio, in which a locally generated transcript
was supplied to a language model; sampled video frames supplied to a
vision-language model; and direct audio processing with compatible
multimodal models. These examples demonstrate implementation
flexibility; they do not prescribe a fixed inference architecture.

Collectively, the evaluations establish that OASIS can support
self-hosted inference for text and selected multimodal workloads. They
are capability evaluations, not claims of model equivalence or
modality-independent performance. Results depend on the model
architecture, serving runtime, context window, preprocessing pipeline,
generation parameters, and information available in the input
representation. OASIS records provider and model identifiers with each
result, while comparative analyses specify the relevant preprocessing
and presentation configuration.

Deployment locality is a property of the complete processing chain, not
only the primary model endpoint. A fully local configuration executes
transcription, primary grading, schema normalization, and any
model-assisted post-processing locally. A configuration that combines a
self-hosted primary model with a hosted downstream service is therefore
reported as hybrid. The evaluated MAPLES workflow that paired a
vLLM-served primary model with a hosted schema-conversion step is one
example, not a required architecture.

\subsection{Standalone, Interactive, and Agent-Native
Operation}\label{standalone-interactive-and-agent-native-operation}

OASIS extends the institutional stack with a standalone execution layer
intended to reduce adoption friction. A user can point the CLI at a
local directory of files and a rubric, inspect a cost estimate, grade a
small sample, and then scale to a full run without bringing up databases
or containerized services. This progressive workflow is technically
important because it narrows failure radius: rubric problems, provider
misconfiguration, or unsupported modalities can be discovered before a
full cohort run.

The guided entry point, \texttt{oasis\ run\ -\/-interactive}, recasts
the same pipeline as a stepwise workflow: environment check, model
selection, smart scan, data review, rubric validation, plan, sample,
grade, and review. The \texttt{-\/-smart-scan} path uses an LLM to
classify messy input directories, infer encounter boundaries, and
surface ambiguous file groupings before grading begins. The CLI-tier
Wayfinder agent applies canonical workflow recipes with infrastructure
fallback gates; the recipes themselves are detailed under Agent-Native
Command Surface.

\begin{longtable}[]{@{}
  >{\raggedright\arraybackslash}p{(\linewidth - 2\tabcolsep) * \real{0.4286}}
  >{\raggedright\arraybackslash}p{(\linewidth - 2\tabcolsep) * \real{0.5714}}@{}}
\toprule\noalign{}
\begin{minipage}[b]{\linewidth}\raggedright
Capability
\end{minipage} & \begin{minipage}[b]{\linewidth}\raggedright
Why it matters
\end{minipage} \\
\midrule\noalign{}
\endhead
\bottomrule\noalign{}
\endlastfoot
Standalone CLI execution & Eliminates infrastructure requirements for
initial evaluation and small-scale grading \\
Interactive terminal UI & Gives users a local surface for browsing runs,
inspecting artifacts, and reviewing outputs \\
MCP tool surface & Makes the same workflow operable by AI assistants and
autonomous agents \\
Shared command-service layer & Keeps CLI, TUI, and MCP behavior aligned
instead of maintaining separate implementations \\
Content-addressable cache & Avoids repeated LLM calls for unchanged
inputs and rubrics \\
Audit and provenance capture & Preserves prompts, routing, outputs, and
review state for later inspection \\
Distributed batch commands & Allows grading work to be split across
multiple machines \\
\texttt{review-session} loop & Turns completed runs into explicit
rubric-improvement and re-grade feedback cycles \\
Transcript-first audio workflow & Lets audio inputs participate in
lower-cost text-capable grading paths after transcription \\
\end{longtable}

The same core command service is exposed through three surfaces:
conventional CLI commands, an interactive full-screen terminal UI, and
an MCP server for agent operation. This parity is deliberate; its
architectural consequences are unpacked in the Agent-Native Command
Surface section.

These capabilities make the standalone path more than a thin wrapper
around provider APIs. Two are developed further below --- distributed
batch, which follows a create/work/merge pattern across machines, and
segment-focused grading of longer media, which reduces cost and sharpens
focus before synthesis. Because these behaviors live in the shared
command-service layer, they reach human operators and agent workflows at
the same time.

\subsection{Progressive Execution and Execution
Plans}\label{progressive-execution-and-execution-plans}

OASIS treats grading as a staged operation rather than as a single
irreversible action. In the current CLI architecture, workflows move
through five checkpoints: \textbf{initialize}, \textbf{setup},
\textbf{dry-run}, \textbf{single}, and \textbf{scale}. In user-facing
terms this appears as \texttt{doctor} followed by plan generation, a
sample run, and only then full grading or distributed batch execution.

\begin{longtable}[]{@{}
  >{\raggedright\arraybackslash}p{(\linewidth - 4\tabcolsep) * \real{0.1795}}
  >{\raggedright\arraybackslash}p{(\linewidth - 4\tabcolsep) * \real{0.2308}}
  >{\raggedright\arraybackslash}p{(\linewidth - 4\tabcolsep) * \real{0.5897}}@{}}
\toprule\noalign{}
\begin{minipage}[b]{\linewidth}\raggedright
Stage
\end{minipage} & \begin{minipage}[b]{\linewidth}\raggedright
Purpose
\end{minipage} & \begin{minipage}[b]{\linewidth}\raggedright
Typical OASIS surface
\end{minipage} \\
\midrule\noalign{}
\endhead
\bottomrule\noalign{}
\endlastfoot
Initialize & Verify provider credentials, environment, and service
reachability & \texttt{oasis\ doctor} \\
Setup & Scan input data, parse rubric, validate mappings and modalities
& \texttt{oasis\ auto-grade} plan mode \\
Dry-run & Compose prompts, estimate work, and surface warnings before
spend & plan output / execution plan \\
Single & Grade one or a few encounters and inspect quality &
\texttt{-\/-sample\ N} \\
Scale & Run the full cohort locally, orchestrated, or via distributed
batch & \texttt{-\/-standalone}, \texttt{-\/-execute},
\texttt{batch\ *} \\
\end{longtable}

This design makes cost and structural validity visible before expensive
calls. A malformed rubric, a missing provider key, or an unsupported
modality can therefore stop the workflow at plan time or sample time
instead of after a full cohort has already consumed tokens.

\begin{figure}

\centering{

\pandocbounded{\includegraphics[keepaspectratio]{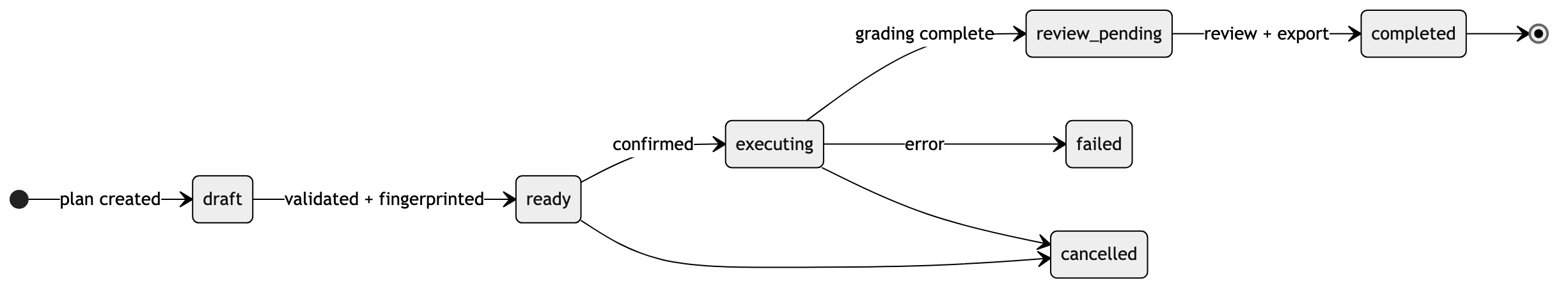}}

}

\caption{\label{fig-execution-plan}Execution-plan lifecycle. Plans are
created as \texttt{draft}, become \texttt{ready} once inputs are
validated and fingerprinted, and move through \texttt{executing} to
\texttt{review\_pending} when grading completes; review and export
transition the plan to \texttt{completed}. Errors mark the plan
\texttt{failed}, and operators can cancel active plans.}

\end{figure}

The progressive pipeline is itself externalized as an artifact. OASIS
writes \texttt{execution-plan.v2} documents as Markdown plus YAML
frontmatter, with immutable step IDs, fingerprinted inputs, waiver
metadata, and an explicit lifecycle (Figure~\ref{fig-execution-plan}).
For a full grading workflow, the canonical step graph is
\texttt{doctor}, \texttt{validate}, \texttt{manifest}, \texttt{rubric},
\texttt{mapping}, \texttt{context}, \texttt{storage}, \texttt{plan},
\texttt{confirm}, \texttt{sample}, \texttt{grade}, \texttt{verify},
\texttt{review}, and \texttt{export}. This is more than user-interface
scaffolding. It gives humans and agents an execution state they can
inspect, annotate, resume, fork, verify, or abandon explicitly.

\subsection{Agent-Native Command
Surface}\label{agent-native-command-surface}

The MCP surface is not generated by wrapping a separate REST API or by
shelling out to CLI subprocesses. Instead, the CLI, interactive terminal
UI, and MCP server all share the same internal \texttt{CommandService}
execution layer. They therefore inherit the same command dispatch,
envelope format, compound operations, and error semantics.

The generated MCP reference exposes the OASIS command surface over
JSON-RPC. Beyond argument schemas, each tool carries planning metadata:
\texttt{x\_oasis\_mutating}, \texttt{x\_oasis\_category},
\texttt{x\_oasis\_prerequisites}, \texttt{x\_oasis\_follow\_up}, and
\texttt{x\_oasis\_typical\_duration}. These annotations are
operationally important because grading workflows mix read-only
discovery with irreversible actions such as uploading files, triggering
pipelines, or mutating plans.

The tool catalog spans discovery, stack management, Elephant operations,
data handling, rubric operations, grouping, pipeline execution,
provenance, review, batch, summary, workflow, and plan-management
families. It also includes compound tools such as \texttt{stack.check},
\texttt{pipeline.grade\_and\_wait}, and \texttt{batch.execute}, plus
discovery helpers such as \texttt{catalog.search} and
\texttt{catalog.describe}. Together, these features make the MCP surface
agent-native rather than merely agent-accessible: the tools are shaped
for planning and recovery, not just for remote invocation.

The shared command-service architecture also reduces drift. A capability
added for human CLI use can become available to the interactive UI and
the MCP server without maintaining a second behavior implementation. For
a platform meant to support both expert users and autonomous assistants,
that parity is itself an architectural property.

\textbf{Model tier separation.} OASIS distinguishes between models
suited for \emph{orchestration} (tool use, workflow reasoning,
multi-step planning) and models suited for \emph{grading} (media
perception, rubric-faithful scoring). The unified model catalog tags
each model with capability flags --- \texttt{OrchestratorCapable} for
models that reliably follow multi-tool workflows,
\texttt{MediaGradeCapable} for models that accept native audio or video
input. This prevents a common failure mode in agent-driven assessment:
using a model that is excellent at text grading but unreliable at
multi-step tool orchestration, or vice versa. The system is opinionated
about defaults but not restrictive; users without a preferred
orchestration provider receive an informational warning rather than a
hard block.

\textbf{Canonical workflow recipes.} The CLI-tier Wayfinder includes a
catalog of canonical workflow recipes covering first-time standalone
grading, orchestrated institutional runs, rubric creation and iteration,
run diagnosis, audio/video workflows, and messy-data triage, among
others. Each recipe specifies a tool sequence with fallback gates: if
Docker or Elephant is unavailable, the agent falls back to the
standalone path rather than failing, so agent-driven operation degrades
gracefully instead of blocking on infrastructure prerequisites.

\begin{figure}

\centering{

\pandocbounded{\includegraphics[keepaspectratio]{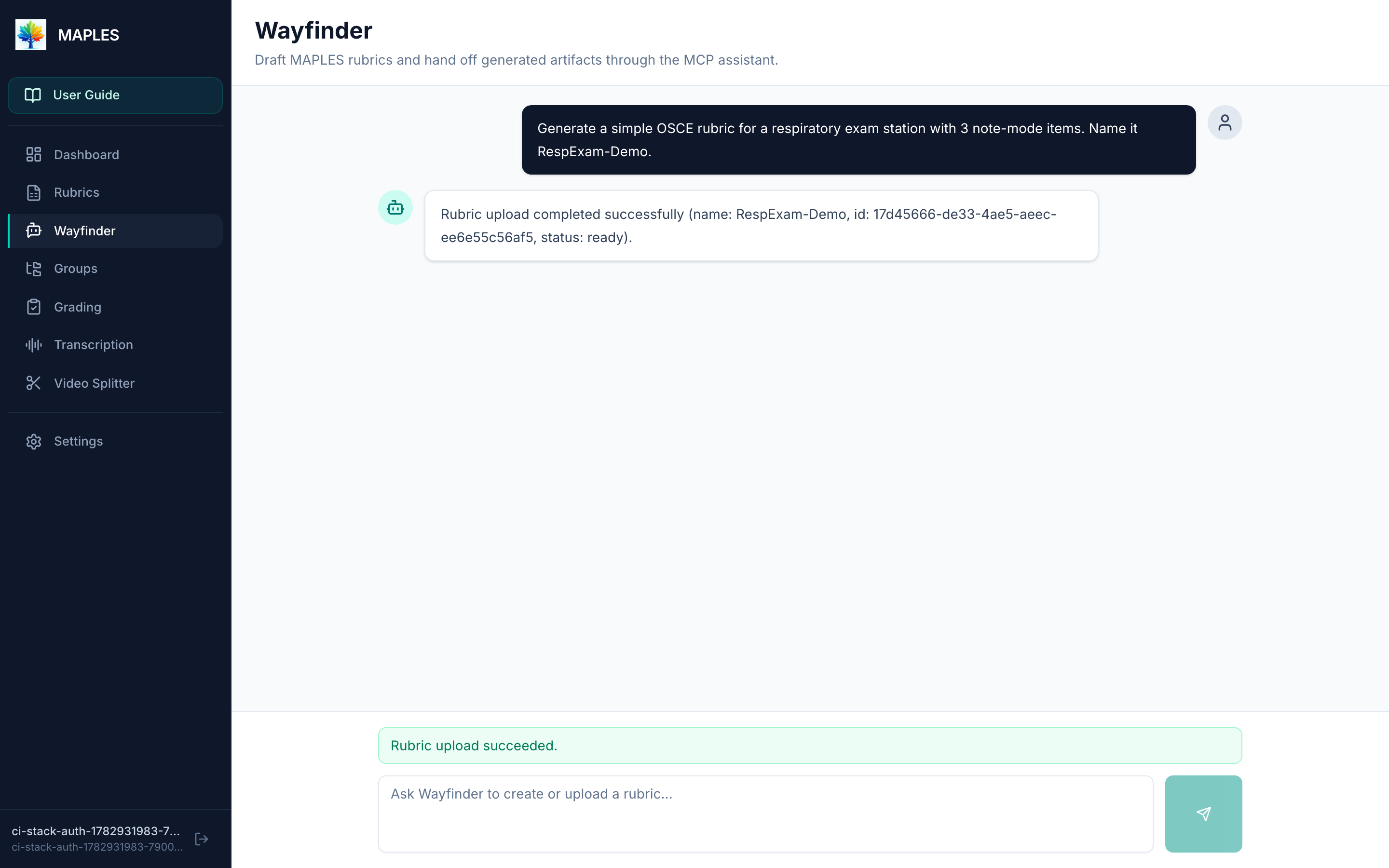}}

}

\caption{\label{fig-wayfinder}The platform-tier Wayfinder assistant in
the MAPLES web interface. A natural-language request produces a
schema-valid rubric that the gateway generates with a versioned
authoring skill and uploads through the authenticated API; the reply is
a deterministic receipt derived from the actual tool result (rubric
name, id, status), and the rubric is immediately available --- versioned
and reviewable --- in the Rubrics view. Captured against a local stack
with synthetic data.}

\end{figure}

\textbf{Platform-tier agent gateway.} The institutional stack carries a
second, distinct MCP server. Where the CLI-tier server wraps the shared
command-service layer, the platform agent gateway proxies the MAPLES
REST API, so web-session agents operate under MAPLES's own
authentication and authorization rather than a parallel permission
model. The trust boundary is deliberate. The gateway itself is
unauthenticated and network-isolated (loopback-bound by default); MAPLES
enforces authentication at its own \texttt{/v1/wayfinder/chat} endpoint
and forwards the signed-in user's token. Agent actions therefore carry
exactly the identity and permissions of the human who asked. For rubric
authoring, the gateway orchestrates versioned skills from the MAPLES
Toolkit submodule (for example, \texttt{generate-rubric-zeroshot} and
its audio, video, and multimodal variants). An evaluator describes the
rubric in natural language; the platform-tier Wayfinder drafts it
against the platform's rubric schema and uploads it through the
authenticated \texttt{upload\_rubric} tool, where it lands as an
ordinary versioned rubric (Figure~\ref{fig-wayfinder}). The exchange in
the figure is enforced, not merely requested: if the model loads a
generation skill and ends its turn without actually calling
\texttt{upload\_rubric}, the gateway forces the tool call, and the reply
shown to the user is a deterministic receipt built from the real API
response --- rubric name, id, and status --- rather than model prose.
The two servers are complementary by construction: one gives agents the
operator's workflow, the other gives agents the institution's workflow,
and both keep human review state authoritative.

\subsection{SimRubrics: Multi-Model Rubric Quality
Assurance}\label{simrubrics-multi-model-rubric-quality-assurance}

The grading pipeline begins before any encounter is processed. The
quality of LLM-based scoring is bounded by the quality of the rubric
itself. A vague criterion such as ``communicates effectively'' produces
unstable judgments regardless of model quality. SimRubrics addresses
this upstream problem by treating rubric authoring as its own technical
workflow rather than as informal prompt writing.

In the standard workflow, a configured LLM provider analyzes each rubric
section. Prompts are template-based: one step extracts the criterion
structure from the uploaded spreadsheet, and another normalizes it into
the OASIS schema with fields such as \texttt{Technique},
\texttt{Purpose}, and \texttt{AdditionalContext}. The result is not a
final score but a cleaner rubric specification for downstream execution.

For higher-stakes rubric development, SimRubrics offers a
\textbf{multi-model consensus} workflow. In Round 1, three frontier
models --- Gemini, Claude, and Grok --- analyze the same rubric item
independently. (SimRubrics' consensus panel is configured separately
from the grading pipeline's providers.) In Round 2, each model critiques
the others' analyses. In Round 3, a synthesis step consolidates the
critiques into a single recommendation describing what to change, why to
change it, and what tradeoffs the change introduces.

The evaluator remains in control throughout. They can accept suggested
changes, modify them, or provide a clarifying directive that triggers
another analysis cycle. Every accepted revision produces an immutable
\texttt{RubricVersion} with JSON and YAML snapshots, so the rubric's
evolution is captured as part of the system record rather than
disappearing into ad hoc spreadsheet edits.

SimRubrics also includes a built-in \textbf{test-grading} feature:
evaluators can upload a sample encounter and grade it against the
current rubric version without invoking the full MAPLES pipeline. This
creates a fast loop for finding brittle criteria before
institutional-scale execution.

The export format is a structured YAML file that mirrors the MAPLES
\texttt{Rubric} Pydantic model:

\begin{Shaded}
\begin{Highlighting}[]
\FunctionTok{name}\KeywordTok{:}\AttributeTok{ }\StringTok{"Fall Assessment — Communication Skills"}
\FunctionTok{stations}\KeywordTok{:}
\AttributeTok{  }\FunctionTok{Observation}\KeywordTok{:}
\AttributeTok{    }\FunctionTok{name}\KeywordTok{:}\AttributeTok{ }\StringTok{"Direct Observation"}
\AttributeTok{    }\FunctionTok{prompts}\KeywordTok{:}
\AttributeTok{      }\FunctionTok{Audio}\KeywordTok{:}
\AttributeTok{        }\KeywordTok{{-}}\AttributeTok{ }\FunctionTok{key}\KeywordTok{:}\AttributeTok{ }\StringTok{"active\_listening"}
\AttributeTok{          }\FunctionTok{system}\KeywordTok{:}\AttributeTok{ }\KeywordTok{[}\StringTok{"..."}\KeywordTok{]}
\AttributeTok{          }\FunctionTok{user}\KeywordTok{:}\AttributeTok{ }\KeywordTok{[}\StringTok{"..."}\KeywordTok{]}
\AttributeTok{          }\FunctionTok{response\_config}\KeywordTok{:}
\AttributeTok{            }\FunctionTok{structured}\KeywordTok{:}\AttributeTok{ }\CharTok{true}
\end{Highlighting}
\end{Shaded}

SimRubrics is optional. For rapid prototyping, users can bypass it and
provide MAPLES with an Excel rubric directly. A third authoring path is
conversational: the platform's embedded Wayfinder assistant can draft a
rubric from a natural-language description using versioned authoring
skills and upload it through the same authenticated API, after which it
is reviewable and replaceable like any other rubric. Whether the rubric
arrives as curated YAML from SimRubrics, as a spreadsheet to be compiled
on upload, or as an agent-drafted artifact, MAPLES ultimately receives
the same normalized rubric representation. The next sections follow that
representation through execution: ingestion, compilation, scoring,
review, and audit.

\subsection{MAPLES: Pipeline Lifecycle}\label{maples-pipeline-lifecycle}

MAPLES is the execution layer that turns a rubric and a collection of
encounters into scored outputs. The six stages below define the outer
lifecycle of a run; the sections that follow unpack the inner mechanics
of the middle stages, where rubric logic becomes prompts, prompts become
typed results, and typed results become reviewed records.

\begin{enumerate}
\def\labelenumi{\arabic{enumi}.}
\tightlist
\item
  \textbf{Group creation.} The user selects encounters from Elephant's
  catalog --- by case, activity, cohort, or date range --- and saves the
  selection as a named group. The resulting manifest is stored in object
  storage.
\item
  \textbf{Rubric ingestion.} MAPLES accepts either an Excel rubric or a
  pre-compiled YAML rubric. In the Excel path, \texttt{rubric\_zipper}
  (the rubric compiler, detailed under The Rubric-to-Prompt
  Transformation) validates the workbook, compiles it into a typed
  \texttt{Rubric} object, and stores the compiled representation.
\item
  \textbf{Pipeline submission.} Validation runs before any LLM calls:
  rubric completeness, provider capabilities against requested
  modalities, and encounter file availability. A run that cannot succeed
  is rejected before tokens are spent.
\item
  \textbf{Per-encounter processing.} A Prefect flow dispatches one task
  per encounter. Encounters run concurrently up to worker capacity.
  Within each encounter, video, audio, and text grading paths run in
  parallel.
\item
  \textbf{Result persistence.} Each scored item is written as a
  \texttt{ResultRow} in PostgreSQL, while aggregated result artifacts
  and rubric snapshots are written to object storage so that the run
  remains inspectable after completion.
\item
  \textbf{Human review.} Evaluators inspect, accept, override, and
  export results in the review interface.
\end{enumerate}

Provider capabilities determine which modality paths a run can use:

\begin{longtable}[]{@{}
  >{\raggedright\arraybackslash}p{(\linewidth - 6\tabcolsep) * \real{0.1786}}
  >{\centering\arraybackslash}p{(\linewidth - 6\tabcolsep) * \real{0.2619}}
  >{\centering\arraybackslash}p{(\linewidth - 6\tabcolsep) * \real{0.2619}}
  >{\centering\arraybackslash}p{(\linewidth - 6\tabcolsep) * \real{0.2976}}@{}}
\toprule\noalign{}
\begin{minipage}[b]{\linewidth}\raggedright
Provider path
\end{minipage} & \begin{minipage}[b]{\linewidth}\centering
Native video grading
\end{minipage} & \begin{minipage}[b]{\linewidth}\centering
Native audio grading
\end{minipage} & \begin{minipage}[b]{\linewidth}\centering
Transcript/text grading
\end{minipage} \\
\midrule\noalign{}
\endhead
\bottomrule\noalign{}
\endlastfoot
Google Gemini & Yes & Yes & Yes \\
OpenAI, Azure OpenAI, or OpenAI-compatible & No & Yes & Yes \\
\end{longtable}

Audio can also route transcript-first --- transcribed once, then graded
as text --- which lets text-only endpoints participate in audio-derived
grading. This capability check is part of the broader execution model:
OASIS attempts to fail early and explicitly when a requested workflow is
structurally invalid. Once a run passes validation, the central
technical step is compilation: the uploaded rubric must be transformed
into prompt batches and typed result schemas before any encounter can be
graded.

\subsection{The Rubric-to-Prompt
Transformation}\label{the-rubric-to-prompt-transformation}

The central abstraction connecting rubric authoring and execution is
that a rubric is treated as a program. The evaluator writes the
specification in a spreadsheet; \texttt{rubric\_zipper} compiles that
specification into prompts and result schemas that drive the grading
pipeline. This compilation step is the contract between rubric design
and runtime behavior.

\textbf{Input format.} Each rubric is an Excel workbook. A
\texttt{\_metadata} sheet stores the rubric name, description, version,
and assessment context as key-value pairs. An optional
\texttt{\_mapping} sheet maps activity/case combinations to station
sheets. A \emph{station} is one scoring context within the assessment
--- for example, one room in an Objective Structured Clinical
Examination (OSCE). Each station sheet contains one row per scorable
item with the following columns:

\begin{longtable}[]{@{}
  >{\raggedright\arraybackslash}p{(\linewidth - 2\tabcolsep) * \real{0.4706}}
  >{\raggedright\arraybackslash}p{(\linewidth - 2\tabcolsep) * \real{0.5294}}@{}}
\toprule\noalign{}
\begin{minipage}[b]{\linewidth}\raggedright
Column
\end{minipage} & \begin{minipage}[b]{\linewidth}\raggedright
Purpose
\end{minipage} \\
\midrule\noalign{}
\endhead
\bottomrule\noalign{}
\endlastfoot
\texttt{ItemKey} & Unique identifier (e.g., \texttt{organization}) \\
\texttt{Mode} & Modality: Audio, Video, or Note (\texttt{Note} is the
rubric format's name for the text modality) \\
\texttt{Section} & Logical grouping within a station \\
\texttt{QuestionText} & What to assess --- the criterion definition \\
\texttt{Response1}--\texttt{Response6} & Scoring descriptors per level;
descriptor \emph{N} corresponds to score \emph{N−1} (e.g., 0--5) \\
\texttt{AdditionalContext} & Optional supplemental grading
instructions \\
\texttt{Technique}, \texttt{Purpose} & Optional fields for video-based
assessment \\
\end{longtable}

The standalone rubric engine also supports a V2 YAML format generated by
\texttt{oasis\ rubric\ init}. V2 rubrics keep item data separate from
prompt text: item keys, scoring levels, modality, and optional context
are stored as structured fields, and the engine composes prompts at
runtime. This allows OASIS to layer domain, program, station, and
evidence context consistently while preserving expert overrides for
items that require hand-authored instructions. V2 rubrics are also
optimized for cost and latency through item batching: ordinary items
with compatible modality and time hints can be grouped into a single
prompt, while items with custom system prompts, user prompts, or
response schemas bypass batching to preserve explicit author intent.
Because compatible items share prompts, item-heavy rubrics see roughly
an order-of-magnitude reduction in LLM calls --- a direct consequence of
batch size rather than a benchmark claim --- without changing the
scoring contract.

\textbf{Compilation.} \texttt{rubric\_zipper} reads the workbook,
validates its structure, groups items by modality and section, and
generates prompt batches. Items within a batch are formatted in a
\texttt{\#\#\#\#}-delimited block:

\begin{verbatim}
####
organization: Is the document organized in a logical, readable
structure with clear sections or headings? Can a reader quickly
find specific information?

SCORING RUBRIC:

0: No discernible structure; information is scattered or
   presented as stream-of-consciousness
1: Some structure is present but inconsistent; key information
   is difficult to locate
2: Well-organized with clear sections, logical flow, and easy
   to scan for specific details.
\end{verbatim}

Multiple items are concatenated into a single prompt. The system
typically batches 8--20 items per prompt to balance API cost against
response complexity.

\textbf{Response schema generation.} For each prompt,
\texttt{rubric\_zipper} dynamically creates a Pydantic model using
\texttt{create\_model()}. Each item key becomes a field whose type is
the modality-specific response model --- \texttt{NoteResponseItem},
\texttt{AudioResponseItem}, or \texttt{VideoResponseItem}. The resulting
schema defines the shape of stored results and is reused by
postprocessing, validation, and provider-native structured-output paths
when available. In this way, the rubric controls both the question being
asked and the structure of the answer the system expects to persist. The
worked example below follows one rubric row through that contract.

\subsection{Worked Example: From Spreadsheet Row to Reviewed
Result}\label{worked-example-from-spreadsheet-row-to-reviewed-result}

The compilation contract is easiest to see by following one rubric item
end to end. Consider the \texttt{organization} item from the
Documentation Quality rubric in the internal synthetic validation
corpus. This item assesses whether a written document is structured
logically, which makes it applicable to incident reports, inspection
summaries, post-encounter notes, or any other structured written
artifact.

\textbf{1. Rubric row (Excel).} \texttt{ItemKey}: \texttt{organization},
\texttt{Mode}: Note, \texttt{Section}: Documentation,
\texttt{QuestionText}: ``Is the document organized in a logical,
readable structure with clear sections or headings? Can a reader quickly
find specific information?'', \texttt{Response1}: ``No discernible
structure; information is scattered or presented as
stream-of-consciousness'', \texttt{Response2}: ``Some structure is
present but inconsistent; key information is difficult to locate'',
\texttt{Response3}: ``Well-organized with clear sections, logical flow,
and easy to scan for specific details.''

\textbf{2. Generated prompt (Stage 1 input; see Two-Stage Grading
below).} The system message is: \emph{``You are an expert evaluator
assessing the quality of a written post-encounter document. Evaluate
based only on the text provided.''} The user message contains the
\texttt{\#\#\#\#}-delimited item block shown above, followed by
instructions to extract evidence from the document, provide a rationale,
and assign a score using the rubric levels.

\textbf{3. Parsed response (Stage 2 output).} In the default two-stage
path, the LLM first returns a free-form response. OASIS then parses that
response into a typed object:

\begin{Shaded}
\begin{Highlighting}[]
\FunctionTok{\{}
  \DataTypeTok{"organization"}\FunctionTok{:} \FunctionTok{\{}
    \DataTypeTok{"evidence"}\FunctionTok{:} \StringTok{"The document uses four labeled headings (Background,}
\StringTok{        Findings, Analysis, Recommendations). Each section contains}
\StringTok{        2{-}3 focused paragraphs. No information appears outside its}
\StringTok{        logical section."}\FunctionTok{,}
    \DataTypeTok{"rationale"}\FunctionTok{:} \StringTok{"Clear hierarchical structure with labeled sections.}
\StringTok{        A reader can locate any category of information without}
\StringTok{        scanning the full document."}\FunctionTok{,}
    \DataTypeTok{"score"}\FunctionTok{:} \DecValTok{2}
  \FunctionTok{\}}
\FunctionTok{\}}
\end{Highlighting}
\end{Shaded}

\textbf{4. Stored provisional result.} The evidence, rationale, and
score are persisted as a \texttt{ResultRow} in PostgreSQL, linked to the
encounter, pipeline run, and rubric version. The result is immediately
inspectable, but it is still provisional: an evaluator can later inspect
it in the review interface, accept it, override it, and export the final
dataset.

This single-item walkthrough scales to the full pipeline. The same
contract is applied to prompt batches rather than one row at a time, and
the same typed result then moves into caching, human review, and the
audit record.

\subsection{Two-Stage Grading}\label{two-stage-grading}

\begin{figure}

\centering{

\pandocbounded{\includegraphics[keepaspectratio]{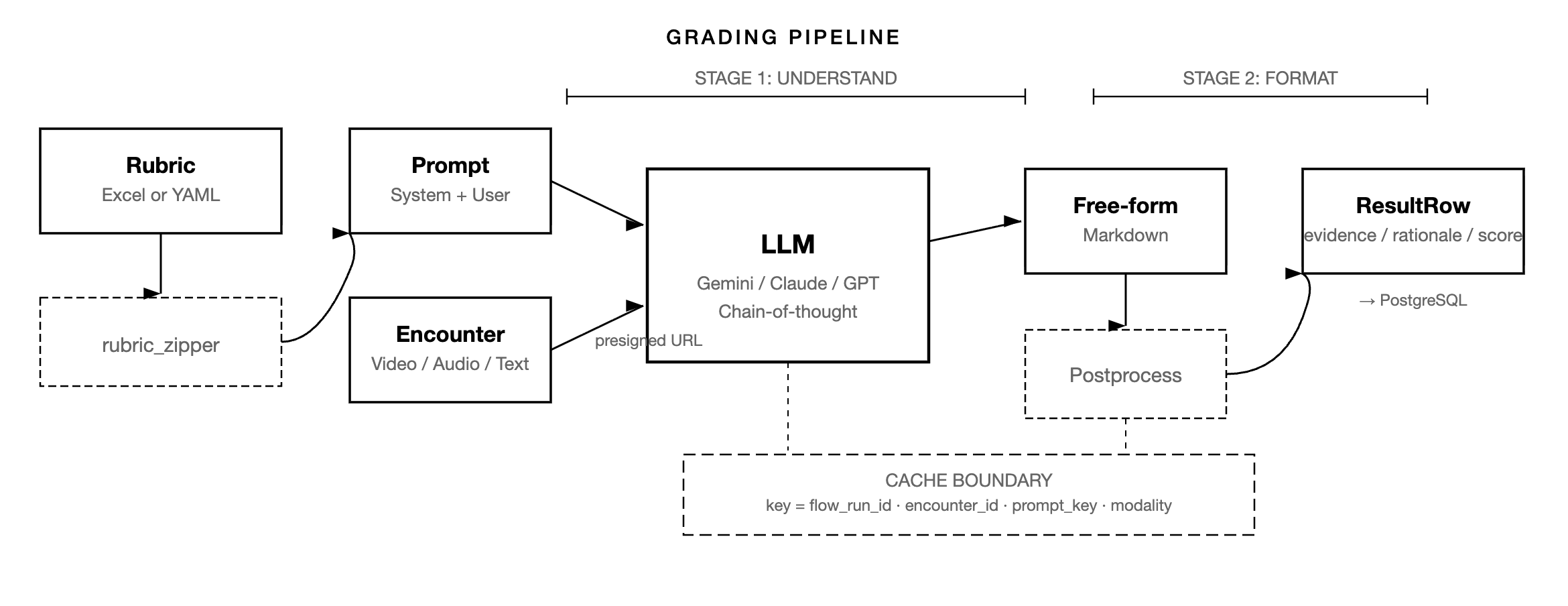}}

}

\caption{\label{fig-pipeline}Grading pipeline. A rubric and encounter
enter from the left. Stage 1 produces a free-form LLM response; Stage 2
parses it into typed fields. Both stages are independently cached with
flow-scoped keys.}

\end{figure}%

The worked example above follows one criterion, but the production path
separates semantic judgment from structural normalization
(Figure~\ref{fig-pipeline}). In the default MAPLES path, \textbf{Stage
1} (\emph{understand}) sends the encounter media (video frames, audio
file, or text) and the rubric prompt to the LLM and receives a free-form
narrative response. \textbf{Stage 2} (\emph{format}) then parses that
response into the typed schema (evidence, rationale, score per item).
Some standalone paths can use provider-native structured output, but the
institutional reference path keeps reasoning and formatting separate.

This separation has three advantages. First, in our experience,
free-form responses produce more detailed reasoning than responses
constrained to a JSON schema during the initial media-analysis step.
Second, if parsing fails, Stage 2 can retry from the recorded Stage 1
output without repeating the expensive media analysis. Third, each stage
is independently cached.

Cache keys are scoped to the pipeline run:
\texttt{\{flow\_run\_id\}-\{encounter\_id\}-\{prompt\_key\}-\{modality\}}.
Including the Prefect \texttt{flow\_run\_id} ensures that retrying a
failed pipeline reuses cached LLM responses from the same run without
cross-contaminating results between runs. The postprocess cache adds a
hash of the input text, so re-parsing only triggers when the raw
response changes.

\subsection{Multimodal Depth: Transcript-Augmented Audio and Segment
Decomposition}\label{multimodal-depth-transcript-augmented-audio-and-segment-decomposition}

The perception-before-assessment principle (Design Decision 3) has
concrete cost and quality consequences for each modality. For text
inputs (\texttt{.docx}, \texttt{.pdf}, \texttt{.rtf}, or plain text),
extraction is automatic and deterministic. For audio and video, OASIS
extracts timestamped transcripts via Gemini Flash and caches them by
file hash. Subsequent grading --- by the same model, a different model,
or on a different machine --- reuses the cached perception artifact
instead of repeating the expensive media call. Audio prompts can then be
graded from transcript text by any text-capable provider, while video
prompts can combine native multimodal perception with transcript
context. This widens provider choice for audio-oriented assessments
without discarding the original media.

The rubric format also supports focused media access. Prompts can carry
a \texttt{time\_hint} indicating that only a particular encounter phase
should influence scoring. For narrowly targeted items, windowed
transcript grading cuts token cost by well over an order of magnitude
relative to full-recording context (internal estimates: roughly 1/45th).
This matters in practice because many performance rubrics assess
phase-specific behaviors rather than the entire encounter uniformly.

For longer recordings, OASIS supports a deeper execution mode
(\texttt{-\/-depth\ deep}) that automatically infers likely time windows
for each rubric item from transcript content. The engine writes a
segment plan explaining why each interval was selected, grades those
segments independently, and then performs a synthesis step that
cross-references evidence across segments. In one internal validation, a
15-minute encounter graded in this mode used roughly 1/40th the tokens
of full-context grading. More importantly, segment decomposition reduces
irrelevant context and makes evidence traces sharper for long multimodal
runs.

\subsection{Content-Addressable Cache and Distributed
Batch}\label{content-addressable-cache-and-distributed-batch}

The standalone grading cache is content-addressable rather than purely
execution-addressable. Its goal is to identify the logical grading unit
independent of any single run identifier, while still distinguishing
exact model-specific outputs. The current design uses two keys:

\begin{Shaded}
\begin{Highlighting}[]
\NormalTok{input\_key  = sha256(rubric\_hash + media\_hash + prompt\_key [+ time\_hint])}
\NormalTok{output\_key = sha256(input\_key + model + provider)}
\end{Highlighting}
\end{Shaded}

The \texttt{input\_key} answers the question ``what are we grading?''
and is model-agnostic. The \texttt{output\_key} answers the question
``what result did this provider/model produce for that input?'' This
separation enables several useful behaviors at once: deduplication for
unchanged inputs, exact cache hits for reruns with the same model,
side-by-side comparison across models for the same content, and portable
provenance records that still make sense outside a particular server run
ID. When the input has been graded by a different model than the one
currently requested, the cache reports the partial hit, enabling the
orchestrator to decide whether to reuse the existing result or re-grade
with the requested provider.

To continue the worked example: the \texttt{organization} item's
\texttt{input\_key} hashes the Documentation Quality rubric snapshot,
the note's bytes, and the prompt key \texttt{organization}. Grading the
same note with a second model leaves that \texttt{input\_key} untouched
and adds a second \texttt{output\_key}, so the two models' scores are
directly comparable as readings of the same input --- and a rerun with
the original model is a cache hit, not a second API call.

The same identity layer supports distributed batch execution.
\texttt{batch\ create} builds a manifest with one work unit per
\texttt{(encounter,\ prompt)} pair, carrying content hashes and rubric
identity. \texttt{batch\ work} allows multiple agents or machines to
process the same manifest concurrently while consulting the shared cache
before each unit. If another machine has already completed that exact
\texttt{(rubric,\ file,\ prompt,\ model)} combination, the work is
skipped immediately rather than recomputed. \texttt{batch\ merge} then
combines agent outputs into a unified result set. In effect, the cache
is not just a local optimization; it is also the coordination substrate
that makes incremental multi-machine scaling safe without requiring a
central queue or lock service.

\subsection{Review Workflow}\label{review-workflow}

Review is not an afterthought; it is the state transition that turns
provisional machine output into reportable assessment data. After
grading completes, every scored item carries the LLM's rationale --- the
evidence it extracted and the reasoning behind its score. The review
interface lets evaluators work through results efficiently:

\begin{itemize}
\tightlist
\item
  \textbf{Filter} by case, modality, score range, learner, or review
  status
\item
  \textbf{Bulk-accept} AI scores for high-confidence items (e.g., all
  items where AI and rubric descriptors align unambiguously)
\item
  \textbf{Override} individual scores with a comment explaining the
  disagreement
\item
  \textbf{Export} the final dataset to Excel, with encounter files
  fetched via Elephant presigned URLs
\end{itemize}

No scores are considered final until an evaluator has reviewed them. The
system tracks three states per item: \texttt{pending},
\texttt{accepted}, and \texttt{reviewed} (the machine score was
overridden by a human). The worked example's \texttt{organization} score
of 2 enters this queue as \texttt{pending}; whether it is accepted
unchanged or overridden with a comment becomes part of the permanent
record. Per-learner aggregated scores and percentiles are computed from
the reviewed results and stored in an aggregation table for downstream
analysis. These review transitions are themselves persisted, so the
final dataset retains both the machine suggestion and the human
adjudication history.

The standalone path carries the same principle into local workflows
through \texttt{review-session}. That command reads the grading trace,
summarizes score distributions, flags outliers or rubric-score
mismatches, and writes a \texttt{session-review.v0} artifact linked back
to the source run. Review therefore becomes part of the improvement loop
rather than a separate spreadsheet clean-up step: inspect evidence,
identify weak rubric items or model blind spots, edit the rubric,
re-grade, and compare the new run through the provenance ledger.

\subsection{Provenance and Audit}\label{provenance-and-audit}

That review history is one part of a broader provenance chain. One of
the distinctive features of the current OASIS platform is that it treats
grading artifacts as first-class technical outputs, not incidental logs.
A run produces more than scores:

\begin{longtable}[]{@{}
  >{\raggedright\arraybackslash}p{(\linewidth - 2\tabcolsep) * \real{0.6087}}
  >{\raggedright\arraybackslash}p{(\linewidth - 2\tabcolsep) * \real{0.3913}}@{}}
\toprule\noalign{}
\begin{minipage}[b]{\linewidth}\raggedright
Run artifact
\end{minipage} & \begin{minipage}[b]{\linewidth}\raggedright
Purpose
\end{minipage} \\
\midrule\noalign{}
\endhead
\bottomrule\noalign{}
\endlastfoot
Run envelope & Identity and configuration of the execution \\
Input manifest & Exactly which encounters and files were graded \\
Rubric snapshot & The immutable rubric version in force \\
Composed prompt batches & The instructions actually sent to the model \\
Provider/model identifiers & Attribution of every output to an exact
model \\
Model outputs & Raw and intermediate responses, pre-parsing \\
Parsed results & Typed scores, evidence, and rationales \\
Evidence indexes & Links from each criterion back to its source
material \\
Replay instructions & How to re-execute the same grading unit \\
Review state & The human adjudication history per item \\
\end{longtable}

Standalone execution writes these artifacts to a local run directory;
institutional deployment links them to run identifiers and stores them
alongside pipeline outputs.

Threaded end to end, the worked example's single spreadsheet row is now
traceable through every artifact in this table: the rubric snapshot
holds its descriptors, a prompt batch holds its compiled question, the
parsed results hold its typed score and evidence, its cache keys give
the grade a stable identity, and the review state holds its human
adjudication --- with replay instructions sufficient to re-execute the
unit. That is the paper's thesis in miniature: one rubric item,
inspectable at every stage of its life.

The traceable unit is therefore not just a final score, but the lineage
from rubric version to prompt batch to model output to review decision.
This provenance model supports practical questions that matter in
high-stakes assessment: What exact input did the model see? Which
provider and model produced a score? Which evidence was extracted for a
given criterion? Was the final value accepted unchanged or overridden by
a reviewer? Because OASIS records these artifacts systematically,
post-hoc inspection and troubleshooting are possible without
reconstructing the workflow from memory or ad hoc log statements.

The audit model also supports product hardening. OASIS can capture audit
sessions spanning CLI commands, MCP tool calls, interactive CLI-tier
Wayfinder conversations, LLM calls, retries, and recovery behavior.
Operators can list, inspect, and analyze these sessions through the
\texttt{audit} command family, while \texttt{inspect}, \texttt{ledger},
and replay/retrace bundles connect local run artifacts back to input
identity. Human operators can inspect runs after failure, but
agent-driven execution benefits just as much: tool invocations, provider
selection, and recovery behavior can all be studied from recorded
artifacts. In this sense, provenance is not only a compliance feature;
it is also an engineering mechanism for making both the human and agent
surfaces more reliable over time.

\subsection{Elephant: Contract-First
Design}\label{elephant-contract-first-design}

If MAPLES is the execution layer, Elephant is the storage and transport
layer that keeps encounters, files, and run inputs addressable over
time. Its design goal is straightforward: grading infrastructure should
depend on a typed contract, not on ad hoc file passing or manual schema
coordination.

Elephant is built API-first. A single \texttt{openapi.yaml}
specification generates both the Go server (via \texttt{oapi-codegen})
and the Python SDK (via \texttt{openapi-python-client}). The contract
between MAPLES and Elephant is enforced at code-generation time in both
languages --- a field added to the spec appears in both the server
handler and the client model. In Go, a missing field is a compile error.
In Python, it is caught by static type checking (mypy). Either way,
contract violations surface before deployment, not at runtime.

Primary keys are UUIDv7: time-ordered, globally unique, and generated
without a central counter. This enables distributed ingestion ---
multiple data sources can upload encounters concurrently without
coordination.

Elephant never proxies file bytes. When MAPLES needs a video for
grading, it requests a presigned URL from Elephant and downloads
directly from MinIO or GCS. A 100 MB video file flows from object
storage to the LLM client; Elephant handles only the URL. This keeps the
API server lightweight regardless of media size.

The data model centers on encounters. A \texttt{learner} is linked to
one or more \texttt{encounters}; each encounter belongs to a
\texttt{case} and an \texttt{activity} and carries one or more
\texttt{encounter\_files} (video, audio, text, transcript, metadata). An
\texttt{encounter\_events} table logs every lifecycle event with a JSONB
payload, forming a complete audit trail. All database queries are
written in raw SQL and compiled to type-safe Go at build time via
\texttt{sqlc} --- no ORM reflection at runtime.

Encounters are deduplicated by a deterministic \texttt{encounter\_key}
(SHA-256 of the encounter's identifying attributes). Files are
deduplicated by content hash. Together, these make ingestion idempotent
--- uploading the same data twice is a no-op, even under concurrent
load. Complex ingestion operations are implemented as multi-step common
table expressions with PostgreSQL advisory locking, so that concurrent
uploads to the same encounter are serialized at the database level
rather than in application code.

Access is controlled by API key roles: \texttt{read\_only},
\texttt{read\_write}, \texttt{read\_write\_delete}, and \texttt{admin}.
MAPLES needs only \texttt{read\_only} access to fetch encounter files
for grading; ingestion scripts use \texttt{read\_write}; administrative
operations (hard deletes, key management) require \texttt{admin}.

A Python SDK (\texttt{elephant-sdk}) provides typed access from MAPLES
and ingestion scripts. The SDK is generated from the same OpenAPI
specification as the server and vendored into the MAPLES tree, so the
integrated stack builds without access to private package registries:

\begin{Shaded}
\begin{Highlighting}[]
\ImportTok{from}\NormalTok{ elephant }\ImportTok{import}\NormalTok{ ElephantClient}

\NormalTok{client }\OperatorTok{=}\NormalTok{ ElephantClient(base\_url}\OperatorTok{=}\StringTok{"http://localhost:8080"}\NormalTok{, api\_key}\OperatorTok{=}\StringTok{"..."}\NormalTok{)}
\NormalTok{encounters }\OperatorTok{=}\NormalTok{ client.search\_encounters(case\_name}\OperatorTok{=}\StringTok{"2025{-}Fall{-}Assessment"}\NormalTok{)}
\NormalTok{url }\OperatorTok{=}\NormalTok{ client.get\_presigned\_url(file\_id}\OperatorTok{=}\StringTok{"018e..."}\NormalTok{)}
\end{Highlighting}
\end{Shaded}

\section{Availability}\label{availability}

Project information and this technical report are available at
\url{https://github.com/JamiesonLabUTSW/oasis} and
\url{https://jamiesonlabutsw.github.io/oasis/}. Application and
component source code, binaries, installation materials, sample data,
and a tagged software release are not included in the current public
publication.

The configurations below characterize the system evaluated for this
report; they are not an assertion that the software can currently be
cloned or installed from the public repository. OASIS was evaluated in
three configurations:

\begin{longtable}[]{@{}
  >{\raggedright\arraybackslash}p{(\linewidth - 4\tabcolsep) * \real{0.2250}}
  >{\raggedright\arraybackslash}p{(\linewidth - 4\tabcolsep) * \real{0.4500}}
  >{\raggedright\arraybackslash}p{(\linewidth - 4\tabcolsep) * \real{0.3250}}@{}}
\toprule\noalign{}
\begin{minipage}[b]{\linewidth}\raggedright
Profile
\end{minipage} & \begin{minipage}[b]{\linewidth}\raggedright
Included surface
\end{minipage} & \begin{minipage}[b]{\linewidth}\raggedright
Primary use
\end{minipage} \\
\midrule\noalign{}
\endhead
\bottomrule\noalign{}
\endlastfoot
\textbf{Standalone} & CLI + local artifacts & Evaluate rubric behavior,
provider compatibility, modality support, and cost with no service
dependencies beyond an API key \\
\textbf{Integrated local stack} & Elephant + MAPLES + review + storage &
Exercise data management, orchestration, and adjudication workflows
under loopback containerized services \\
\textbf{Agent-native (optional)} & MCP servers (CLI tier and optional
platform gateway) + audit bundle & Evaluate assistant and autonomous
operation against the same workflows used by humans \\
\end{longtable}

In the evaluated implementation, the standalone profile ran on a laptop
with one researcher. The integrated local profile used one Docker host
and exercised Elephant, MAPLES, storage, orchestration, and review
together. The Wayfinder gateway was a separate opt-in profile. Shared or
hosted operation requires a separately governed deployment adapter and
is not the reference configuration documented here.

The same rubric and artifact model supports a local single-user run, a
small institutional pilot, or a managed production service. Throughput
in production is bounded primarily by LLM provider rate limits rather
than by the orchestration layer itself; encounters are processed
concurrently up to the provider's token-per-minute quota. In one
internal synthetic run using Gemini 2.5 Pro, observed cost was roughly
\$0.01 and elapsed time about 17 seconds. Those values are illustrative,
provider-dependent, and not a reproducibility guarantee; native media
grading varies substantially with provider and recording length.

\textbf{Data residency.} What leaves the local environment depends on
the complete pipeline configuration, not only the primary model. When
transcription, primary grading, schema conversion, and post-processing
all run locally, text grading---and transcript-first audio grading---can
operate without sending encounter content to a hosted model provider. A
local primary paired with any hosted downstream stage is instead a
hybrid workflow; that hosted stage receives its configured input, which
can include primary-model text, evidence excerpts, or other
source-derived content. Native video and hosted-audio grading transmit
media to the configured provider (currently Gemini for the supported
native-video path). Perception artifacts, caches, and run records remain
in the configured deployment storage.

The private implementation workspace uses unit tests, HTTP integration
tests, scenario packs, end-to-end harnesses, MCP coverage checks, and CI
gates. These controls characterize the verification discipline discussed
below; they are not public release artifacts in the current publication.
A later software release may include a versioned synthetic data
inventory and its attribution record under separately stated terms.

\subsection{Responsible Use}\label{responsible-use}

OASIS is an assessment-support system, not an autonomous decision-maker.
AI-generated scores are provisional by design: the review workflow
exists so that a qualified evaluator examines, accepts, or overrides
machine output before it is treated as final. Ultimate responsibility
for any judgment, grade, or decision made with the tool rests with the
expert and institution using it, not with the software or the models it
calls. Deployments remain subject to the operating institution's own
requirements for privacy, consent, data retention, and assessment
governance. Release materials exclude learner data and institutional
assessment artifacts; public examples are synthetic.

\subsection{Verification and Release
Discipline}\label{verification-and-release-discipline}

OASIS treats verification as part of the system surface. The core
workflow-harness design is organized as five layers: contract/schema
checks (L0), atomic execution (L1), behavioral contracts (L2), workflow
scenarios (L3), and resilience/recovery (L4). Around that spine, the
current integrated workspace adds curated auto-grade scenario suites,
direct MCP coverage tests for the generated tool catalog, battle-test
scripts, and real agent-scenario tests. A separate
\texttt{mission\ canary} command provides fail-closed readiness
validation for low-cost preflight checks, including policy scoring and
optional plan evidence before larger grading work proceeds.

At the private implementation state reviewed for this report, the
integrated workspace's automation ran as a 12-job GitHub Actions
pipeline. Nine non-LLM jobs covered CI preflight linting, Go
vet/race/snapshot coverage, shared binary builds, Python contract and
docs-drift checks, workflow-harness smoke tests, VHS terminal
regression, MAPLES unit tests, scenario validations, and readiness
gates. A sequential integration job exercised the Docker stack plus the
shell end-to-end harness. Two push-only LLM jobs added direct Gemini
execution and agent-smoke scenarios with audit capture, while a separate
nightly workflow exercised the MAPLES-mediated Gemini path end to end.
This structure matters because it makes regressions in grading,
contracts, tool metadata, or audit behavior visible during development
rather than leaving them latent in production.

For an assessment platform, that verification posture is part of the
engineering claim. The same product that records provenance for learner
scoring also records enough evidence to debug its own command surfaces,
workflow harnesses, and release process.

\section{Operational Experience}\label{operational-experience}

OASIS has been in production at UT Southwestern Medical Center since
Fall 2023, grading clinical skills exams (OSCEs) across multiple
programs.

\begin{longtable}[]{@{}
  >{\raggedright\arraybackslash}p{(\linewidth - 2\tabcolsep) * \real{0.5625}}
  >{\raggedright\arraybackslash}p{(\linewidth - 2\tabcolsep) * \real{0.4375}}@{}}
\toprule\noalign{}
\begin{minipage}[b]{\linewidth}\raggedright
Measure
\end{minipage} & \begin{minipage}[b]{\linewidth}\raggedright
Value
\end{minipage} \\
\midrule\noalign{}
\endhead
\bottomrule\noalign{}
\endlastfoot
In production since & Fall 2023 \\
Encounters processed & \textgreater{} 7,000 \\
Learners covered & \textgreater{} 3,000 \\
Modalities & text, audio, video \\
Item-level percent agreement with human graders & 93--96\%
(chance-uncorrected) \\
AI--human Cohen's κ (recent multimodal deployment) & 0.830 \\
Human--human inter-rater κ (same setting) & 0.732 \\
Estimated reduction in manual grading effort & 95--97\% \\
\end{longtable}

The agreement and effort figures are operational observations rather
than controlled measurements. The percent-agreement range summarizes
chance-uncorrected item-level agreement observed across production
cohorts and modalities; the effort estimate reflects the shift from
item-by-item human scoring to review-first adjudication --- bulk
acceptance plus targeted overrides --- for equivalent cohorts. The κ
comparison comes from a companion study of one recent multimodal
deployment (Kang et al. 2026).

These figures should be interpreted as real-world deployment
observations, not a universal benchmark across tasks or institutions.
They nevertheless show that the platform is operationally viable at
meaningful scale. The same machinery also serves as research
instrumentation: reproduction studies of published grading benchmarks
have been executed end-to-end through the CLI, with each experimental
cell --- a model, station, modality, and encounter-subset configuration
--- declared as a spec, executed against the standalone or institutional
path, and joined to evaluator scores through the run artifacts. Just as
important, the system reflects lessons from that deployment history:
fail-fast validation, typed result persistence, explicit review state,
and provenance capture were added because operational assessment
requires them. Published studies using OASIS include work in NEJM AI
(Jamieson et al. 2024), JMIR AI (Kang et al. 2025), and Discover
Artificial Intelligence (Campbell et al. 2025).

The experiment surface has also been used to compare hosted and local
open-weight models against common frozen inputs. Internal engineering
campaigns have exercised vLLM-served Gemma and Qwen models for direct
text grading, local-ASR-to-text audio grading, frame-presented video
grading, and selected native-audio paths. The main lesson is
methodological as much as operational: model identity cannot be
separated from presentation identity. A sparse-frame video path and a
native continuous-video path are different experimental conditions, just
as transcript-first audio and native waveform input are different
conditions. OASIS preserves those distinctions so that an apparent model
difference is not silently presented as a model-only comparison.

\section{Limitations}\label{limitations}

The current limitations fall into three categories: provider dependence,
assessment-design dependence, and reproducibility boundaries.

\textbf{Provider dependence and scope.} Open-weight text grading and
transcript-first audio are established execution paths, but multimodal
capability remains model-, runtime-, and presentation-dependent. Video
grading is currently best served by Google's Gemini, which remains the
supported provider for native continuous-video input. Open-weight
vision-language models can receive extracted frames, but that
representation is not equivalent to native video and should not be
compared without a matched-presentation design. Selected open-weight
models can accept native audio, but support depends on model
architecture, serving-runtime version, context limits, and available
accelerator hardware. The architecture is not Gemini-specific---the
provider factory is designed so that any compatible model can be
integrated---but the practical quality frontier remains
provider-dependent by modality. OASIS is also a batch system for
post-hoc grading of recorded encounters; it does not attempt real-time
scoring during a live assessment.

\textbf{Assessment-design dependence.} OASIS faithfully executes the
rubric it is given. Ambiguous criteria, overlapping score levels, or
weakly defined constructs produce unstable scores, and prompt wording
can shift score distributions even when authors intend to assess the
same construct. SimRubrics mitigates this by making rubric refinement
explicit, but it does not eliminate the need for expert rubric design.
Likewise, OASIS produces scored outputs and rationales, not full
psychometric interpretation; item response theory (IRT),
generalizability theory (G-theory), and related analyses remain
downstream tasks.

\textbf{Reproducibility boundaries.} OASIS records rubric versions,
prompts, provider identifiers, outputs, and review decisions, but it
cannot prevent a hosted model provider from changing behavior behind a
stable API name. Open-weight operation improves the ability to retain an
exact model artifact, but model weights alone are insufficient:
tokenizer or processor versions, serving runtime, quantization or data
type, generation configuration, and media presentation can all affect
the result. Exact score reproducibility therefore depends on binding
those runtime details in addition to OASIS's own run provenance.
Complete runtime identity remains an area for continued schema and
release-harness work.

\section{Conclusion}\label{conclusion}

Rubric-based assessment with LLMs is often demonstrated one prompt at a
time; deploying it is a systems problem. OASIS's answer is an
architecture in which the rubric is a compiled specification, execution
is progressive and cost-visible, grading identity is
content-addressable, human review is a first-class state transition, and
every run leaves an auditable trail from rubric version to final
adjudicated score. The canonical workflow runs at two altitudes: a
single binary on a laptop and a containerized local Elephant + MAPLES
stack. Hosted APIs and self-hosted open-weight models participate
through the same rubric, review, comparison, and provenance structure,
making local inference a first-class operating mode rather than a
separate experimental fork. Optional typed agent surfaces, including the
Wayfinder assistant, extend those paths without becoming onboarding
requirements. Shared hosting remains a separately governed site adapter.
Adoption can therefore grow from a one-afternoon evaluation to
cohort-scale operations without changing assessment semantics. One
property of the agent design generalizes beyond assessment: because
agents act through the same authenticated interfaces and review gates as
the people they assist, automated operation carries the same identity,
permissions, and accountability as human operation.

Nearly three years of production use at UT Southwestern, spanning more
than 7,000 encounters, grounds the design in operational reality: the
platform's contracts, gates, and provenance mechanisms exist because
assessment at scale demanded them. This report documents that
operational experience and the resulting architecture. A later software
release may provide code, sample data, and verification harnesses under
separately stated terms. OASIS is offered in that spirit --- an
assessment system in which every score can be traced, reviewed, and
replayed.

\section{Acknowledgements}\label{acknowledgements}

This work was supported by the Office of the President, UT Southwestern
Medical Center. AI tools (Claude, Codex, Copilot, Gemini) were used
during development and documentation; all technical claims and
manuscript text were reviewed by the authors, who take responsibility
for the final content.

\section*{References}\label{references}
\addcontentsline{toc}{section}{References}

\phantomsection\label{refs}

\end{document}